\documentclass[prl,aps,twocolumn]{revtex4}

\usepackage{graphicx}
\usepackage{amsmath}

\begin{document}

\title{Composite domain walls in flat nanomagnets: the magnetostatic limit}

\author{H. Youk}

\author{G.-W. Chern}

\author{K. Merit}

\author {B. Oppenheimer}

\author{O. Tchernyshyov}

\affiliation{Department of Physics and Astronomy, 
The Johns Hopkins University, 
3400 N. Charles St., 
Baltimore, Maryland 21218}

\begin{abstract}
We discuss the structure of the so-called ``vortex'' domain walls 
in soft magnetic nanoparticles.  A wall of this kind is a composite
object consisting of three elementary topological defects: two 
edge defects with winding numbers $-1/2$ and a vortex with a winding 
number $+1$ between them.  We provide a qualitative model accounting
for the energetics of such a domain wall.  
\end{abstract}

\maketitle

Magnetic nanoparticles generate considerable interest as prospective
building blocks for nonvolatile random-access memory \cite{Zhu00}.
Storing of information bits is made possible by the existence of two
(or more) stable magnetic configurations.  Switching between the two
stable states can be achieved by applying an external magnetic field
or by injecting current.  In either case the switching process
proceeds through the formation of complex transient patterns of
magnetization \cite{FQZhu04}.  Building fast and reliable magnetic
memory thus requires a thorough understanding of magnetization
dynamics in these nanomagnets.

In magnetic nanoparticles with the geometry of strips and rings the
switching creates domains in which the magnetization is forced by the
magnetostatic forces to be parallel to the edge.  These domains grow
and shrink at the expense of one another until one of them occupies
the entire sample \cite{FQZhu04,Klaeui03}.  This process can also be
viewed as the {\em creation, propagation, and annihilation of domain
walls.}  Thus the question of local stability and dynamics of the
magnetic configurations can be answered by studying the static and
dynamic properties of domain walls.

Domain walls in submicron rings and strips have a considerably complex
structure.  For instance, McMichael and Donahue \cite{McMichael97}
have observed, among others, configurations termed ``transverse'' and
``vortex'' domain walls.  In a previous paper \cite{OT05} we pointed
out that the transverse walls are composite objects built from two
elementary topological defects.  In the limit where exchange
interaction is the dominant force (thin and narrow strips or rings),
the two elementary defects are vortices with fractional winding
numbers $+1/2$ and $-1/2$; these defects are confined to the edges
because of their fractional topological charges.  To our knowledge,
these edge defects were first discussed by Moser \cite{Moser04} and
Kurzke \cite{Kurzke04}.

In this paper we analyze the structure of domain walls in a strip in a
different limit where the dominant forces are magnetostatic.  This
limit, in which both the thickness and width of a strip exceed the
exchange length $\lambda = \sqrt{A/\mu_0 M_0^2}$ \cite{Hubert}, is
relevant to the ongoing experimental studes \cite{FQZhu04,Klaeui03}.
The nonlocal nature of the dipolar interactions \cite{Hubert} makes
the analysis considerably more difficult.  Valuable information
concerning the global structure of a domain wall is provided by
topological considerations \cite{OT05}.  Under very general
circumstances, a domain wall in a nanostrip is a composite object
containing several elementary topological defects, some of which
reside in the bulk and others at the edge.  The topology restricts
possible compositions of a domain wall thus providing a basis for
selecting appropriate trial states.  

In a companion paper \cite{MMM1} we have identified three elementary
topological defects that survive the transition from the exchange
limit to the magnetostatic regime: the vortex (winding number $n =
+1$), the antivortex ($n = -1$) and one of the edge defects ($n =
-1/2$).  Bare-bones versions of these defects can be constructed using
van den Berg's method \cite{Berg86} wherein the exchange energy is
initially neglected and the magnetostatic energy is minimized
absolutely by preventing the appearance of magnetic charge $-\nabla
\cdot \mathbf{M}$.  (The meat can be grown by including the exchange
interaction perturbatively.)  Only the vortex retains its original
shape; the antivortex morphs into a cross tie (two intersecting
90-degree Neel walls); the $-1/2$ defect looks like a cross tie pinned
at the edge.  The $+1/2$ edge defect likely has a high magnetostatic
energy.  We have also shown \cite{MMM1} that the simplest domain wall
in this limit is expected to contain two $-1/2$ edge defects and a
vortex.  We next discuss the structure of a domain wall in this limit.

\begin{figure}
\includegraphics[width=0.99\columnwidth]{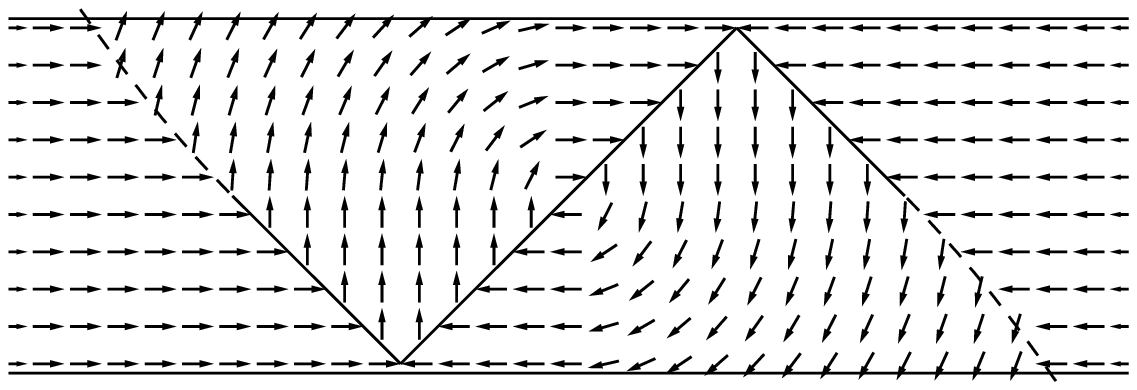}
\includegraphics[angle=90,width=0.99\columnwidth]{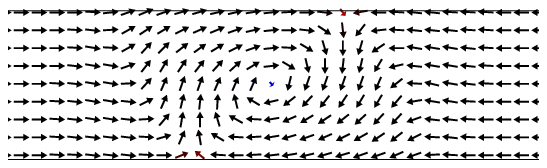}
\caption{Top: a magnetization configuration free of bulk magnetic
charges, $-\nabla \cdot \mathbf{M} = 0$, and containing two $-1/2$
edge defects and a $+1$ vortex in the middle.  Parabolic segments
of Neel walls are shown by dashed lines.  Bottom: a head-to-head
vortex wall obtained in a micromagnetic simulation using OOMMF
\cite{oommf} in a permalloy strip of width $w = 500$ nm and thickness
$t = 20$ nm.  }
\label{fig-center}
\end{figure}

A head-to-head domain wall carries a fixed nonzero amount of magnetic
charge ($2 M_0 t w$ in a strip of width $w$), with a finite density in
the bulk, $-\nabla \cdot \mathbf{M} \neq 0$, or at the film edge,
$\hat{\mathbf{n}} \cdot \mathbf{M} \neq 0$ (or, most likely, both).
Therefore van den Berg's method is not, strictly speaking, applicable.
Nonetheless, an examination of the detailed structure of a vortex wall
(bottom panel of Fig.~\ref{fig-center}) \cite{McMichael97} shows that
it indeed contains two $-1/2$ edge defects and a $+1$ vortex in the
middle.  The edge defects share one of their Neel walls; the vortex
resides at its middle point.

In what follows we consider a model of the vortex domain wall that is
free of {\em bulk} magnetic charge.  Thus all of the charge $2 M_0 t
w$ is expelled to the edges.  Under this restriction it is possible to
construct a vortex domain wall by piecing together the vortex and two
$-1/2$ edge defects as shown in Fig.~\ref{fig-center}.  The resulting
structure contains domains with uniform and curling magnetization.  In
a strip $|y|<w/2$ with the shared Neel wall $x=y$ and the vortex core
at $(v,v)$, the two curling domains in the regions $\pm v < \pm y <
w/2$ are separated by parabolic Neel walls $(x-v)^2 = (2y \pm w)(2v
\pm w)$ from domains with horizontal magnetization; they also merge
seamlessly with other uniform domains along the lines $x = v$ and $y =
v$.

This trial state exaggerates the accumulation of surface magnetic
charge thereby overestimating the normal component of magnetization
$\hat{\mathbf{n}} \cdot \mathbf{M}$ at the edge.  Nonetheless it
captures the major features of a vortex domain wall.

\begin{figure}
\includegraphics[width=0.99\columnwidth]{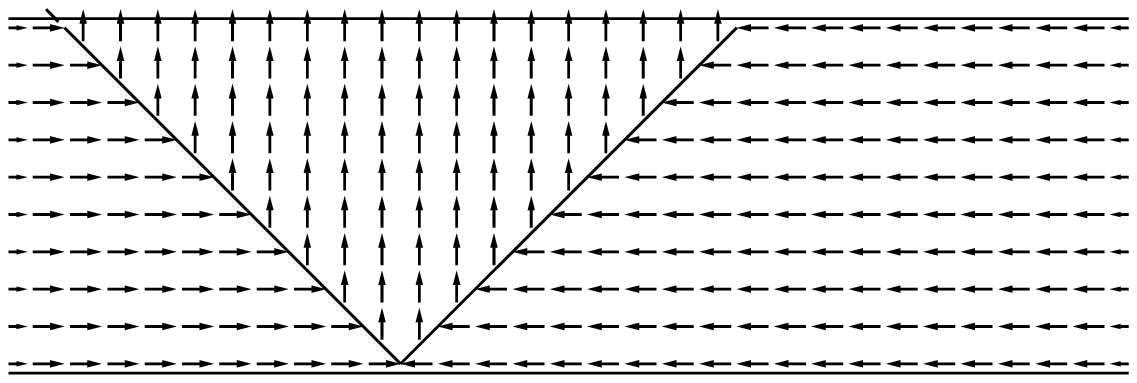}
\includegraphics[angle=-90,width=0.99\columnwidth]{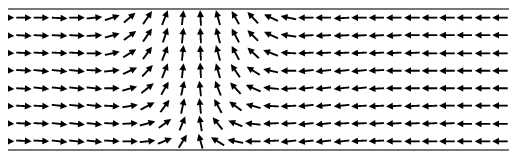}
\caption{Top: a model vortex wall with the vortex absorbed by the
edge.  Bottom: a transverse wall observed in a numerical simulation
(permalloy, $w = 80$ nm, $t = 20$ nm.)}
\label{fig-off}
\end{figure}

The construction of a domain wall out of the three defects is not
unique and has (at least) one degree of freedom: the vortex can be
placed anywhere along the shared Neel wall $x=y$.  When the vortex
core reaches the edge, it is absorbed by the $-1/2$ edge defect.
Their fusion creates an edge defect with the winding number $+1 - 1/2
= +1/2$ (Fig.~\ref{fig-off}).  As can be seen from the figure, the
$+1/2$ defect is rather extended (length $2w$) and, in accordance with
our previous remarks, contains magnetic charge (all of it at the edge
in this model).  This structure is topologically equivalent to the
transverse domain wall in the exchange limit \cite{OT05}, where both
edge defects are pointlike.  The transverse walls observed in
experiments \cite{Klaeui03} and simulations \cite{McMichael97} are
midway between these two extremes: the $+1/2$ defect definitely has a
wider core, although its extent is less than $2w$.

\begin{figure}
\includegraphics[width=0.99\columnwidth]{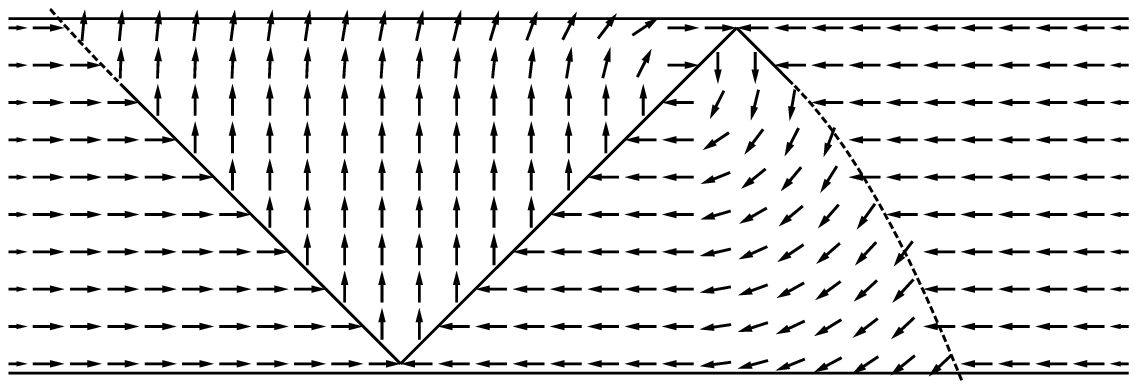}
\includegraphics[angle=-90,width=0.99\columnwidth]{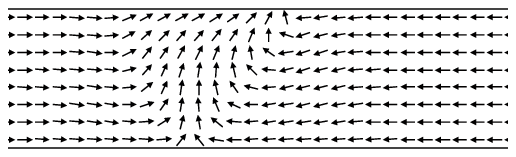}
\caption{Top: a model vortex wall with the vortex near the edge.
Bottom: a similar configuration observed in a numerical simulation
(permalloy, $w = 200$ nm, $t = 20$ nm.)}
\label{fig-edge}
\end{figure}

To determine the equilibrium configuration of the composite wall we
computed the total energy of the composite wall and minimized it with
respect to the vertical coordinate $v$ of the vortex.  The
energy is the sum of the following terms.

\begin{subequations}

The magnetostatic energy coming from the Coulomb-like interaction of
the magnetic charges spread along the edges with the line densities
$\lambda_{1,2} = t \, \mathbf{M} \cdot \hat{\mathbf{n}}_{1,2} = \pm t
M_0 \sin{\theta}$ for the upper and lower edges, respectively.  It
includes the interaction of magnetic charges on the same edge and on
different edges:
\begin{eqnarray}
E_{ii} &=& \frac{\mu_0}{8\pi} 
\int \frac{\lambda_i(x) \, \lambda_i(x')}{|x-x'|} \, dx \, dx',
\nonumber\\
E_{ij} &=& \frac{\mu_0}{8\pi} 
\int \frac{\lambda_i(x) \, \lambda_j(x')}{\sqrt{w^2 + (x-x')^2}} \, dx \, dx',
\label{eq-ms}
\end{eqnarray}
The total magnetostatic energy $\sum_{i=1}^2 \sum_{j=1}^2 E_{ij}$ is
of the order $A w (t^2/\lambda^2) \log{(w/t)}$.  It represents the dominant
contribution in sufficiently wide and thick strips.

The energy of the Neel walls can be computed as a line integral
\begin{equation}
E_{\mathrm{walls}} = t \int \sigma(\ell) \, d\ell,
\label{eq-walls}
\end{equation}
where $d\ell$ is a line element of the wall.  The wall surface tension
$\sigma$ depends on the angle of rotation across the wall, which stays
at 90 degrees along straight segments and varies along parabolic ones.
See Ref.~\onlinecite{MMM1} for details of the calculation.  This term
is of the order $Atw/\lambda$.

The exchange energy 
\begin{equation}
E_{\mathrm{exchange}} = A t \int_\Omega (\nabla \theta)^2 \, dx \, dy,
\end{equation}
where $\Omega$ is the area around the vortex where magnetization
curles.  This term is of the order $A t \log{(w/\lambda)}$.

\end{subequations}

\begin{figure}
\includegraphics[width=\columnwidth]{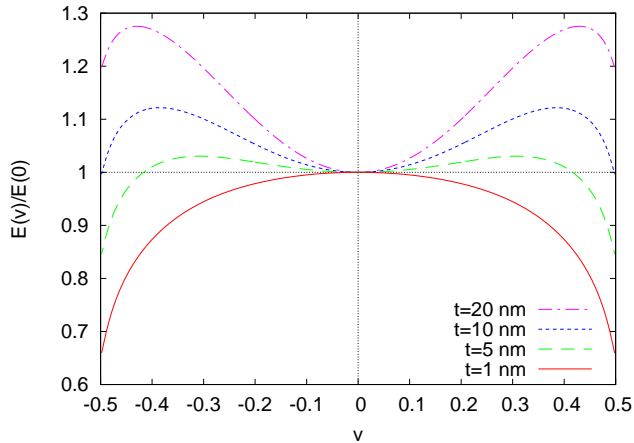}
\caption{Energy of the vortex domain wall as a function of the vortex
position $v$ at a fixed strip width $w = 50$ nm for several thicknesses
$t$.}
\label{fig-curve1}
\end{figure}

An investigation of the phase diagram and a quantitative comparison 
of this model with the numerical and experimental results is currently 
under way.  In what follows we report some preliminary findings.  

The evolution of the energy curve $E(v)$ at a fixed width $w$
and varying thickness $t$ is shown in Fig.~\ref{fig-curve1}.  For
substantially wide and thick strips, the one and only minimum of
energy is achieved with the vortex in the middle of the strip, in
agreement with numerical simulations \cite{McMichael97}.

As the cross section decreases, a local minimum develops with the
vortex core at the edge of the strip.  In this configuration
(Fig.\ref{fig-off}) the vortex and the $-1/2$ edge defect have merged
to form an extended $+1/2$ edge defect.  The configuration is highly
reminiscent of the transverse wall \cite{McMichael97} that is known to
coexist with the vortex wall over a range of cross sections
\cite{Klaeui03}.  The transverse wall becomes a global minimum of
energy when the cross section becomes small enough.  The vortex wall
remains locally stable until it becomes a local maximum of energy.

(The reader should note that the curve for thickness $t = 1$ nm shown
in Fig.~\ref{fig-curve1} is only an extrapolation: the energetics of
the Neel wall in very thin films is a nonlocal problem \cite{Hubert}
to which our estimate based on Eq.~(4) in the companion paper
\cite{MMM1} does not apply.  Nonetheless, the overall trend reflected
by the shape of the curve is correct: in the exchange limit domain
walls containing 3 defects are locally unstable.)

In addition to these two wall configurations, which have been
previously discussed in the literature, we have found two metastable
states that corrsepond to local minima of energy $E(v)$.  One of them
has the vortex core rather close to but not exactly at the edge (top
panel of Fig.~\ref{fig-edge}).  We have observed domain walls of this
kind in numerical simulations (bottom panel of Fig.~\ref{fig-edge}).
The other metastable state occurs when the energy curve $E(v)$ has two
symmetric minima around $v = 0$.  This is a vortex wall with the
vortex core slightly off center.  Because the off-center minima are
rather shallow it may be difficult to observe such states in practice:
even slight imperfections of the nanoparticle can change the potential
landscape $E(v)$.

The simple model of a domain wall in the magnetostatic limit presented
in this paper shows a qualitative agreement with observations.  Its
quantitative comparison with available experimental and numerical data
is in progress.  At a minimum, the model provides an insight into the
nature of the vortex domain walls in a regime relevant to experiments.
It corroborates our earlier suggestion \cite{OT05} that domain walls
in nanostrips can be viewed as composite objects, which may be helpful
in understanding their dynamical properties.  

{\bf Acknowledgment.}  We thank C.-L. Chien, J.-G. Zhu, and F. Q. Zhu
for discussions.  This work was supported in part by the NSF Grant
No. DMR05-20491.

\bibliography{micromagnetics}

\end{document}